\magnification=1200
\baselineskip=18pt
\tolerance=100000
\overfullrule=0pt

\centerline{\bf ANGULAR MOMENTUM, QUATERNION, OCTONION, AND}
\centerline{\bf LIE-SUPER ALGEBRA OSP(1,2)$^*$}
\bigskip
\centerline{by}
\centerline{Susumu Okubo}
\bigskip
\centerline{Department of Physics and Astronomy}
\centerline{University of Rochester, Rochester, NY 14627 U.S.A.}

\vskip 2truein

\noindent {\bf \underbar{Abstract}}

\medskip

We will derive both quaternion and octonion algebras as the Clebsch-Gordan 
algebras based upon the su(2) Lie algebra by considering angular momentum 
spaces of spin one and three. If we consider both spin 1 and ${1 \over 2}$ 
states, then the same method will lead to the Lie-super algebra osp(1,2). 
Also, the quantum generalization of the method is discussed on the basis of 
the quantum group su$_{\rm q}$(2).

\vskip 2truein

\baselineskip=13pt

\item{*}Paper dedicated to the 90th birthday of Professor Ta-You Wu. 
It will be published in the Proceedings of the Commemoration of Prof. Wu
 by World Scientific Publications.

\vfill\eject

\baselineskip=18pt

\noindent {\bf 1. \underbar{Introduction}}

\medskip

The Clebsch-Gordan coefficients of the angular momentum algebra in the 
quantum mechanics have a rich structure.  We have shown in ref. 1 (referred 
to as I hereafter) that the Clebsch-Gordan recoupling of the spin 3 system 
will effectively lead to the octonion algebra. Let $\psi_j (m) 
\equiv | j, m>$ with $m = j, j-1, \dots, - j$ be the 
standard eigenfunction for the angular momentum $j$ with
 $j = 0, {1 \over 2}, 1, {3 \over 2} , \dots$.  Let 
$j$ be a integer and introduce now a $2 j +1$ dimensional algebra with the 
product defined by
$$\psi_j (m_1) \cdot \psi_j (m_2) = b_j \sum_{m_3} C  \pmatrix{j &j &j \cr
\noalign{\vskip 4pt}%
m_1 &m_2 &m_3\cr}
 \psi_j (m_3)\eqno(1.1)$$
for a constant $b_j$. Here,  $C  \pmatrix{j_1 &j_2 &j_3 \cr
\noalign{\vskip 4pt}%
m_1 &m_2 &m_3\cr}$ is the Clebsch-Gordan coefficient for recoupling of two 
angular momentum states $j_1$ and $j_2$ into $j_3$. Note that we are 
restricting here for consideration only of $j_1 = j_2 = j_3 = j$. Since we 
have
$$    C  \pmatrix{j_1 &j_2 &j_3 \cr
\noalign{\vskip 4pt}%
m_1 &m_2 &m_3\cr} = \delta_{m_1 + m_2, m_3} 
C  \pmatrix{j_1 &j_2 &j_3 \cr
\noalign{\vskip 4pt}%
m_1 &m_2 &m_3\cr} \eqno(1.2a)$$
as well as
$$C  \pmatrix{j_1 &j_2 &j_3 \cr
\noalign{\vskip 4pt}%
m_1 &m_2 &m_3\cr} = (-1)^{j_1 +j_2 -j_3} 
C  \pmatrix{j_2 &j_1 &j_3 \cr
\noalign{\vskip 4pt}%
m_2 &m_1 &m_3\cr} \quad ,           
\eqno(1.2b)$$
\smallskip
\noindent we may rewrite Eq. (1.1) as
$$\psi_j (m_1) \cdot \psi_j (m_2) = b_j C_j (m_1 , m_2) 
\psi_j (m_1 + m_2) \eqno(1.3)$$
which satisfies
$$\psi_j (m_1) \cdot \psi_j(m_2) =
(-1)^j \psi_j (m_2) \cdot \psi_j (m_1) \quad ,\eqno(1.4)$$
where we have set
$$C_j (m_1 , m_2) = 
C  \pmatrix{j &j &j \cr
\noalign{\vskip 4pt}%
m_1 &m_2 &m_1 + m_2\cr} \eqno(1.5)$$
for simplicity. Especially, Eq. (1.4) implies that the algebra is 
commutative for $j$ = even but anti-commutative for $j$ = odd. We may also 
introduce a bilinear (not sesqui-linear) form by
$$<\psi_j(m_1)|\psi_j(m_2)>\  = (2j +1)^{1 \over 2} C  \pmatrix{j &j &0 \cr
\noalign{\vskip 4pt}%
m_1 &m_2 &0\cr} = (-1)^{j-m_1} \delta_{m_1 + m_2, 0} \eqno(1.6)$$
now for both integer as well as half-integer values of $j$.  Then, we will 
have
$$<\psi_j (m_2)|\psi_j (m_1)>\  = (-1)^{2j} <\psi_j (m_1)|\psi_j (m_2)>
\eqno(1.7)$$
which is symmetric for integer $j$ but anti-symmetric for half integer $j$. 
 It also satisfies the associative trace condition for integer $j$, i.e.
$$<\psi_j (m_1)\cdot \psi_j (m_2)|
\psi_j (m_3)>\  =\  <\psi_j (m_1)|\psi_j (m_2) \cdot 
\psi_j (m_3)>
\eqno(1.8)$$
 as we may easily verify from  properties of the Clebsch-Gordan 
coefficient. We have shown in I that the algebra defined by Eq. (1.3) give 
the Lie algebra su(2) for $j=1$ and the 7-dimensional exceptional Malcev 
algebra$^{2)}$ for the case of $j=3$. They are intimately connected with the 
quaternion and octonion algebras as follows. We will adjoin the spin 0 
state $e_0 \equiv |j=0,m=0>$ and modify Eq. (1.3) as
$$\psi_j(m_1)* \psi_j(m_2) = a_j<\psi_j (m_1)|\psi_j(m_2)>e_0 +
b_j C_j (m_1, m_2) \psi_j (m_1 +m_2) \eqno(1.9a)$$
where $e_0$ acts as the unit element, i.e.
$$\eqalignno{e_0* \psi_j (m) &= \psi_j (m) *e_0 = \psi_j (m) \quad , &(1.9b)\cr
e_0*e_0 &= e_0 \quad . &(1.9c)\cr}$$
We also set
$$\eqalignno{<e_0|\psi_j(m)> &=\  < \psi_j(m)|e_0>\  = 0 \quad, &(1.10a)\cr
<e_0|e_0> &= 1 \quad . &(1.10b)\cr}$$
If we choose the constant $a_j$ in Eq. (1.9a) suitably, then the modified 
algebra now lead to the quaternion and octonion algebras, respectively for 
$j=1$ and $j=3$. The group-theoretical reasons behind these statements will 
be found in I.
In this note, we will first explicitly demonstrate these facts in terms of 
identities among Clebsch-Gordan coefficients in section 2.  If we consider  
a system consisting of $j=1$ and $j = {1 \over 2}$, then the resulting 
algebra turns out to be the Lie-super algebra$^{3)}$ osp(1,2).  Similarly, 
we can generalize the octonion algebra into a super algebra by  now using  
spins 0, 3, and ${3 \over 2}$. Finally, we will make a comment on the 
possibility of constructing quantum quaternion and quantum octonion 
algebras by use of the quantum Clebsch-Gordan coefficients$^{4)}$ of the 
quantum group su$_{\rm q}$(2).

\medskip

\noindent {\bf 2. \underbar{Quaternion and Octonion Algebras}}

\medskip

Let us first consider the simplest case of spin one. It is then easy to 
verify that the Clebsch-Gordan coefficients satisfy the special relation of 
form
$$C_1(m_1,m_2)C_1(m_1+m_2,m_3) = {1 \over 2} \{ (-1)^{m_2}
\delta_{m_2 +m_3,0} - (-1)^{m_1} \delta_{m_1+m_3,0}\} \eqno(2.1)$$
for any values 0, and $\pm 1$ for $m_1,\ m_2$, and $m_3$. As the 
consequence, the product defined by Eq. (1.3) satisfies
$$\eqalign{(\psi_1 (m_1)\cdot \psi_1 (m_2))\cdot \psi_1 (m_3) =
&{1 \over 2} (b_1)^2 \{ <\psi_1 (m_1)|\psi_1(m_3)> \psi_1(m_2)\cr
& - <\psi_1 (m_2)|\psi_1(m_3)> \psi_1(m_1)\} \quad .\cr}\eqno(2.2)$$
Writing 3 generic elements of the algebra as
$$\eqalignno{x &= \sum^1_{m=-1} \alpha (m) \psi_1 (m) &(2.3a)\cr
y &= \sum^1_{m=-1} \beta (m) \psi_1 (m) &(2.3b)\cr
z &= \sum^1_{m=-1} \gamma (m) \psi_1 (m) &(2.3c)\cr}$$
for constants $\alpha (m)$, $\beta (m)$, and $\gamma (m)$, then we have 
$$\eqalignno{x \cdot y &= -y \cdot x \quad , &(2.4a)\cr
(x \cdot y) \cdot z &= {1 \over 2} (b_1)^2 \{ <x|z>y\  - <y|z>x\} &(2.4b)
\cr}$$
when we multiply $\alpha (m_1) \beta (m_2) \gamma (m_3)$ to both sides of 
Eq. (2.2) and sum over $m_1,\ m_2,\ {\rm and}\ m_3$. Cyclically 
interchanging $x \rightarrow y \rightarrow z \rightarrow x$, and adding 
all, it leads to
$$(x \cdot y)\cdot z + (y \cdot z )\cdot x + (z \cdot x) \cdot y = 0 
\eqno(2.5)$$
where we used the symmetry condition $<x|y>\  =\  <y|x>$ because of Eq. (1.7).  
Together with $x \cdot y = - y \cdot x$ by Eq. (2.4a), this implies that 
the present algebra is indeed a Lie algebra. It is easy to see that it is 
isomorphic to the su(2) for $b_1 \not= 0$ when we identify
$$\psi_1 (0) = - {1 \over \sqrt{2}} b_1 J_3 \quad , \quad \psi_1 (\pm 1) = 
{i b_1 \over 2} J_\pm \quad , \eqno(2.6)$$
and write $x \cdot y = [ x,y]$ which then leads to the familiar relation 
$[J_3, J_\pm] = \pm J_\pm$, and $[J_+, J_-] = 2 J_3$.

Let us next adjoin the unit element $e_0$ and consider the modified algebra 
Eq. (1.9) with
$$x*y = a_1 <x|y>e_0 + x \cdot y \quad . \eqno(2.7)$$
We calculate then
$$\eqalign{(x*y)*z &= a_1 <x|y>z + a_1 <x \cdot y| z> e_0 +
(x \cdot y) \cdot z\cr
x*(y*z) &= a_1 <y|z>x + a_1 <x | y \cdot z> e_0 +
x \cdot (y \cdot z) \cr}$$
so that
$$(x*y)*z - x*(y*z) = a_1 <x|y>z - a_1 <y|z>x
+ (x \cdot y)\cdot z - x \cdot (y\cdot z) \eqno(2.8)$$ 
where we used $<x \cdot y|z>\  =\  <x |y \cdot z>$ by Eq. (1.8). However
$$\eqalign{(x \cdot y) \cdot z - x \cdot (y \cdot z) &= (x \cdot y) 
\cdot z + ( y \cdot z) \cdot x = - (z \cdot x) \cdot y \cr
&= - {1 \over 2} (b_1)^2 \{ <z |y>x\  - <x |y>z \}\cr}$$
from Eqs. (2.4) and (2.5).  Eq. (2.8) then leads to 
$$(x*y)*z - x*(y*z) = \bigg[ a_1 + {1 \over 2} (b_1)^2 \bigg] \{
<x|y>z\  - <y|z>x\} 
\quad .$$
Therefore, if we choose the constant $a_1$ to be
$$a_1 = - {1 \over 2} (b_1)^2 \eqno(2.9)$$
we will find
$$(x * y)*z = x*(y*z) \eqno(2.10)$$
so that the new product is associative . It is easy to verify that this 
together with the unit element $e_0$ leads to the quaternion algebra.

We will next consider the case of $j = 3$. Although we will no longer have 
such a simple
 relation as Eq. (2.1), the following identity can be verified to be 
valid:
$$\eqalign{C_3 (m_1, m_3) &C_3 (m_2 ,m_1 + m_3) + C_3
(m_2, m_3)C_3 (m_1 , m_2 + m_3)\cr
&= {1 \over 6} \{ 2 (-1)^{m_1} \delta_{m_1 +m_2 ,0} - 
(-1)^{m_3} [ \delta_{m_1+m_3 ,0} 
+ \delta_{m_2 +m_3 ,0}]\} \cr} \eqno(2.11)$$
for values of $m_1,\ m_2,\ m_3$ being $0,\ \pm 1,\ \pm 2,\ {\rm and}\ 
\pm 3$.  Setting
$$\eqalign{x &= \sum^3_{m=-3} \alpha (m) \psi_3 (m) \quad , \cr
y &= \sum^3_{m=-3} \beta (m) \psi_3 (m) \quad , \cr
z &= \sum^3_{m=-3} \gamma (m) \psi_3 (m) \quad , \cr} \eqno(2.12)$$
Eq. (2.11) leads now to
$$(x \cdot z) \cdot y + (y \cdot z) \cdot x = {1 \over 6}
(b_3)^2 \{ 2< x|y>z\  - <x|z>y\  - <y|z>x\} \eqno(2.13)$$
when we note Eqs. (1.3) and (1.6).  As we remarked in I, Eq. (2.13) 
implies that it corresponds to the 7-dimensional simple Malcev 
algebra.$^{2)}$ Moreover, if we adjoin the unit element $e_0$ as in Eqs.
 (1.9) and (2.7), then it gives the octonion algebra, provided that we assign 
a suitable value of $a_3$. However, we will not go into detail here which 
can be found in I.

The case of $j =2$ may also be  of some interest, since we now
 have $x \cdot y = 
y \cdot x$.  In that case, the Clebsch-Gordan coefficients satisfy
$$\eqalign{C_2 (&m_1,m_2) C_2 (m_3 ,m_1+m_2) \cr
&+ C_2 (m_2 ,m_3)C_2 (m_1,m_2
+m_3)
+ C_2 (m_3,m_1)C_2 (m_2,m_3+m_1)\cr
= &{2 \over 7} \{ (-1)^{m_1} \delta_{m_1+m_2,0} + (-1)^{m_2}
\delta_{m_2+m_3,0} + (-1)^{m_3} \delta_{m_3+m_1,0}\} \cr}\eqno(2.14)$$
instead of Eq. (2.11) for values of $m_1,\ m_2,\ m_3$ being $0,\ \pm1,\ \pm 
2$. We will then have the cubic equation
$$x^3 = {2 \over 7} (b_2)^2 <x|x>x \eqno(2.15)$$
for generic element $x$ where $x^3 = (x \cdot x) \cdot x = x \cdot (x \cdot 
x)$. If we adjoin the unit element $e_0$, then it will give a Jordan 
algebra. However, we will not discuss the details.

It is sometimes more convenient to use quantities associated with the 
Cartesian, rather than polar coordinates. 
For example, the spin one system may be labelled simply as a vector 
$\phi_\mu$ for $\mu = 1,2,3$. Then, the product Eq. (1.1) will be simply 
written as
$$\phi_\mu \cdot \phi_\nu = b_1^\prime \sum^3_{\lambda =1} \epsilon_{\mu \nu 
\lambda} \phi_\lambda \eqno(2.16)$$
for another constant $b^\prime_1$, where $\epsilon_{\mu \nu \lambda}$ is 
the totally anti-symmetric Levi-Civita symbol in 3-dimension. Choosing 
$b^\prime_1 = 1$ and introducing the new
 product now by $\phi_\mu * \phi_\nu = - 
\delta_{\mu \nu} e_0 + \sum^3_{\lambda =1} \epsilon_{\mu \nu \lambda} 
\phi_\lambda$, it immediately gives
 the quaternion algebra.  Similarly, the spin 
3 system can be specified$^{5)}$ by the totally symmetric traceless tensor 
$\phi_{\mu \nu \lambda} (\mu ,\nu , \lambda = 1, 2,3)$ i.e.
$$\eqalign{{\rm (i)} \qquad\qquad\qquad
&\phi_{\mu \nu \lambda} =\ {\rm symmetric\ in}\ \mu,\  \nu, \ 
{\rm and}\ \lambda\cr
{\rm (ii)}\qquad\qquad\qquad 
 &\sum^3_{\mu =1} \phi_{\mu \mu \lambda} = 0 \ \ \ \ (\lambda = 1,2,3
) \quad .\cr} \eqno(2.17)$$
We now introduce the dot product by
$$\eqalign{\phi_{\mu \nu \lambda} \cdot \phi_{\alpha \beta \gamma} = &{b \over
 3!3!} \sum_{P, P^\prime} \sum^3_{\tau =1} \{ \delta_{\mu \alpha}
\epsilon_{\nu \beta \tau} \phi_{\lambda \gamma \tau}\cr
&- {1 \over 5} \delta_{\mu \nu} \epsilon_{\lambda \beta \tau} 
\phi_{\alpha \gamma \tau} + {1 \over 5} \delta_{\alpha \beta} 
\epsilon_{\gamma \nu \tau} \phi_{\mu \lambda \tau}\}\cr}\eqno(2.18)$$
for another constant $b$, where the summations on $P$ and $P^\prime$ stand 
for 3! permutations of $\mu,\ \nu,\ \lambda$, and of $\alpha,\ \beta,\ 
\gamma$, respectively. Choosing $b = -5$, and identifying
$$\eqalign{e_1 &= - \sqrt{{3 \over 2}} \phi_{233} \quad ,\cr
e_3 &= {1 \over \sqrt{10}} (\phi_{222} - 3 \phi_{112} )\quad ,\cr
e_5 &= - \sqrt{{3 \over 5}} (\phi_{311} - \phi_{322}) \quad , \cr
e_7 &= \phi_{333} \quad ,\cr}
\qquad\qquad\quad
\eqalign{e_2 &= 2 \sqrt{{3 \over 5}} \phi_{123} \quad ,\cr
e_4 &= \sqrt{{3 \over 2}} \phi_{133} \quad ,\cr
e_6 &= - {1 \over \sqrt{10}} (\phi_{111} - 3
\phi_{122}) \quad ,\cr}\eqno(2.19)$$
we can verify that Eq. (2.18) is equivalent to
$$e_A \cdot e_B = \sum^7_{C=1} f_{ABC}\  e_C \eqno(2.20)$$

\vfill\eject 

\noindent for $A,B = 1,2 \dots, 7$ where $f_{ABC}$ is totally anti-symmetric 
constants in $A,\ B,\ C$ with values of $0,\ \pm 1$ 
as is tabulated in I.  Then, adding the unit element $e \equiv e_0$, the 
algebra defined by 
$$e_A * e_B = - \delta_{AB} e_0 + \sum^7_{C=1} f_{ABC}\  e_C \eqno(2.21)$$
gives the standard octonion algebra.

\medskip

\noindent {\bf 3. \underbar{Lie-super Algebra OSP(1,2)}}

\medskip

If we consider algebras containing both integer and half-integer spin 
states, it will lead to super-algebras, where the integer spin states 
correspond to bosonic elements while the half-integer ones give the 
fermionic components. As a example, consider the system consisting of $j=1$ 
and $j = {1 \over 2}$, where we would have
$$\eqalignno{\psi_1 (M_1) \cdot \psi_1 (M_2) &= b_1 \sum_{M_3}
C\pmatrix{1&1&1\cr
\noalign{\vskip 4pt}%
M_1 &M_2 &M_3\cr} \psi_1 (M_3) \quad , &(3.1a)\cr
\noalign{\vskip 4pt}%
\psi_{1 \over 2} (m_1) \cdot \psi_{1 \over 2} (m_2) &= b_2 \sum_M
C\pmatrix{{1 \over 2} &{1 \over 2}&1 \cr
\noalign{\vskip 4pt}%
m_1 &m_2 &M\cr} \psi_1 (M) \quad , &(3.1b)\cr 
\noalign{\vskip 4pt}%
\psi_1 (M_1) \cdot \psi_{1 \over 2} (m_1) &= a_1 \sum_{m_2}
C\pmatrix{1&{1 \over 2}&{1 \over 2}\cr
\noalign{\vskip 4pt}%
M_1 &m_1 &m_2\cr} \psi_{1 \over 2} (m_2) \quad , &(3.1c)\cr 
\noalign{\vskip 4pt}%
\psi_{1 \over 2} (m_1) \cdot \psi_1 (M_1) &= a_2 \sum_{m_2}
C\pmatrix{{1 \over 2}&1&{1 \over 2}\cr
\noalign{\vskip 4pt}%
m_1 &M_1 &m_2\cr} \psi_{1 \over 2} (m_2) \quad , &(3.1d)\cr}$$
for some constants $a_j$ and $b_j$. Note that Eqs. (3.1a) and (3.1b) imply
$$\eqalignno{\psi_1 (M_1) \cdot \psi_1 (M_2) &= 
 - \psi_1 (M_2) \cdot
\psi_1 (M_1) &(3.2a)\cr
\psi_{1 \over 2} (m_1) \cdot \psi_{1 \over 2} (m_2) &=  \psi_{1 \over 2}
 (m_2) \cdot
\psi_{1 \over 2} (m_1) &(3.2b)\cr}$$
because of Eq. (1.2b),  while the commutability between $j=1$ and $j = {
1 \over 2}$ components is not determined since the constants $a_1$ and $a_2
$ are arbitrary. However, the symmetry strongly suggests the choice of $a_1 
= a_2$ in Eqs. (3.1c) and (3.1d) so that we have
$$\psi_1 (M_1) \cdot \psi_{1 \over 2} (m_1) = - \psi_{1 \over 2} (m_1) 
\cdot \psi_1 (M_1) \quad . \eqno(3.3)$$ 
Then, assigning the grade of 0 and 1 for $j=1$ and $j = {1 \over 2}$ 
components, respectively, it defines a super-algebra, since two generic 
elements $x$ and $y$ obey  
$$x \cdot y = -(-1)^{xy} y \cdot x \eqno(3.4)$$
in the standard convention where 
$$(-1)^{xy} = \cases{-1\ \ , & if both $x$ and $y$ are fermionic, i.e.
spin ${1 \over 2}$\cr
\noalign{\vskip 4pt}%
+1 \ \ ,& otherwise\cr} \quad . \eqno(3.5)$$   
Moreover, if we choose the value of $a_1 = a_2$ suitably, then it can be 
verified to give a Lie-super algebra with the Jacobi identity
$$(-1)^{xz} (x \cdot y) \cdot z + (-1)^{yx} (y \cdot z) \cdot x + (-1)^{zy}
(z \cdot x)\cdot y = 0 \quad . \eqno(3.6)$$
Here, if we wish, we can use the more familiar notation of $[x,y]$ or $[x,
y\}$ instead of $x \cdot y$. Further, the resulting Lie-super algebra 
corresponds to the ortho-symplectic one osp(1,2).

In order to prove these assertions made above, it is more convenient to use 
quantities in the Cartesian coordinate, where $\phi_\mu (\mu = 1,2,3)$ 
refers to spin 1  and the spinor $\xi_j (j=1,2)$ represents spin ${1 \over 
2}$. Since $b_1$ and $b_2$ in Eqs. (3.1a) and (3.1b) can always be suitably 
renormalized by adopting suitable normalizations for $\psi_1(M)$ and 
$\psi_{1 \over 2} (m)$, the corresponding relations in the cartesian 
coordinate may be rewritten as
$$\eqalignno{\phi_\mu \cdot \phi_\nu &= i \sum^3_{\lambda =1} \epsilon_{\mu
 \nu \lambda} \phi_\lambda \quad , \quad (\mu, \nu = 1,2,3) \quad , &(3.7a)\cr
\phi_\mu \cdot \xi_j &= - \xi_j \cdot \phi_\mu = a^\prime
\sum^2_{k=1} \xi_k (\sigma_\mu)_{kj} \quad , &(3.7b)\cr
\xi_j \cdot \xi_k &= - {i \over 2} \sum^3_{\lambda =1} (\sigma_2 
\sigma_\lambda)_{jk} \phi_\lambda \quad , \quad
(j,k = 1,2) \quad . &(3.7c)\cr}$$
Here, $\sigma_\mu (\mu = 1,2,3)$ are standards $2 \times 2$ Pauli matrices, 
and we note 
$$(\sigma_2 \sigma_\lambda)^T = \sigma_2 \sigma_\lambda \quad , \quad
\sigma_2^T = - \sigma_2$$
for the transposed matrix.
The Jacobi identity Eq. (3.6) can be readily verified from Eqs. (3.7), 
if the constant $a^\prime$ in Eq. (3.7b) is chosen to be $a^\prime = {1 \over
2}$
 which we assume hereafter.  To show next that the Lie-super algebra is
osp(1,2), 
 we first rewrite Eq. (3.7a) by introducing $X_{ab} (a,b = 1,2)$ by

\vfill\eject

$$X_{11} = -2 (\phi_1 + i \phi_2) \quad , \quad  X_{22} = 2 ( \phi_1 - i
\phi_2)$$
\vskip -.2truein
\line{\hfill  (3.8)}
\vskip -.2truein
$$X_{12} = X_{21} = - 2 \phi_3$$
so that Eq. (3.7a) is rewritten as
$$\eqalignno{{\rm (i)}\qquad &X_{ab} = X_{ba} &(3.9a)\cr
{\rm (ii)} \qquad &X_{ab} \cdot X_{cd} = \epsilon_{bc} X_{ad} +
\epsilon_{ac} X_{bd} + \epsilon_{bd} X_{ac} +
\epsilon_{ad} X_{bc} &(3.9b)\cr}$$
for values of $a,\ b,\ c,\ d = 1,2$ where we have set
$$\epsilon_{11} = \epsilon_{22} = 0 \qquad , \qquad \epsilon_{12} =
 - \epsilon_{21} = 1 \eqno(3.10a)$$
and hence 
$$\epsilon_{ab} = - \epsilon_{ba} \equiv (i \sigma_2)_{ab} \ \ ,\ \ (a,b=1,2) 
\quad. \eqno(3.10b)$$
Note that Eqs. (3.9) is the symplectic Lie algebra sp(2) which is isomorphic to
 su(2) by Eq. (3.8). Similarly, by setting
$$u_1 = - 2 \xi_1 \quad , \quad u_2 = 2 \xi_2 \quad , \eqno(3.11)$$
Eqs. (3.7b) is rewritten as 
$$X_{ab} \cdot u_j = -u_j \cdot X_{ab} = \epsilon_{aj} u_b + 
\epsilon_{bj} u_a \eqno(3.12)$$
if we choose $a^\prime = {1 \over 2}$. Finally, Eq. (3.7c) leads to
$$u_j \cdot u_k = X_{jk}\quad , \quad (j,k=1,2)  \quad . \eqno(3.13)$$
Now, we add an extra index 0 in addition to 1 and 2, and set
$$\eqalign{X_{j0} &= X_{0j} = u_j \quad , \quad (j=1,2)\cr
X_{00} &= 0 \quad .  \cr}  \eqno(3.14)$$
Then Eqs. (3.9), (3.12), and (3.13) are rewritten as 
$$\eqalign{X_{AB} \cdot X_{CD} = &\epsilon_{BC} X_{AD} + (-1)^{B \cdot C}
\epsilon_{AC} X_{BD}\cr
& + (-1)^{B \cdot C} \epsilon_{BD} X_{AC} + 
(-1)^{A\cdot (B +C)} \epsilon_{AD} X_{BC}\cr} \eqno(3.15)$$
for $A,\ B,\ C,\ D = 0, 1, \ {\rm and}\ 2$. Here, we have set
$$\epsilon_{AB} = \cases{\epsilon_{ab} &if $A=a$ and $B=b$\cr
\noalign{\vskip 4pt}%
1 &if $A=B=0$\cr
\noalign{\vskip 4pt}%
0 &otherwise \cr}\quad . \eqno(3.16)$$
Especially, both $\epsilon_{AB}$ and $X_{AB}$ satisfies the symmetry 
conditions
$$\eqalignno{\epsilon_{AB} &= - (-1)^{A \cdot B} \epsilon_{BA} \quad ,
&(3.17a)\cr
X_{AB} &= (-1)^{A \cdot B} X_{BA} &(3.17b)\cr}$$
where $(-1)^{A \cdot B}$ is defined by 
$$(-1)^{A \cdot B} = \cases{ -1\ , &if $A=B=0$\cr
\noalign{\vskip 4pt}%
+1 &otherwise\cr} \eqno(3.18)$$
since the index 0 corresponds to the fermionic variable while other ones 1 
and 2 refer to the bosonic ones. The relation Eq. (3.15) with Eqs. (3.17)
defines the Lie-super algebra osp(1,2) if we identify $x \cdot y = [x,y]$. 
In this connection, we simply mention the fact that Lie-super algebra 
osp(n,m) is intimately related to para-statistics$^{6)}$ where the boson and
fermion operators do no longer commute with each other.

We will next introduce a bilinear form by
$$\eqalignno{<\phi_\mu|\phi_\nu> &= \delta_{\mu \nu} \quad , \quad (\mu, 
\nu = 1,2,3) &(3.19a)\cr
<\xi_j | \xi_k> &= i (\sigma_2)_{jk} = \epsilon_{jk} \quad , \quad (j,k=1,2)
&(3.19b)\cr
<\xi_j | \phi_\mu> &=\  <\phi_\mu|\xi_j> \  = 0 \quad . &(3.19c)\cr}$$
Then, it satisfies
$$\eqalignno{{\rm (i)} \quad &<x|y>\  = 0\ ,\ {\rm unless}\ x\ {\rm and}\ 
y\ {\rm are\ both\ bosonic\ or\ fermionic} &(3.20a)\cr
{\rm (ii)}\quad &<y|x>\  = (-1)^{xy}<x|y> &(3.20b)\cr
{\rm (iii)}\quad &<x\cdot y|z>\  =\  <x|y\cdot z> &(3.20c)\cr
{\rm (iv)}\quad &<x|y>\ {\rm is\ non-degenerate}  &(3.20d)\cr}$$
so that $<\cdot | \cdot>$ is a supersymmetric bilinear non-degenerate
associative form.

We now adjoin the unit element $e_0$ and define a new product by
$$\eqalignno{x*y &= -<x|y> e_0 - i x \cdot y &(3.21a)\cr
x*e_0 &= e_0 * x = x \quad , \quad e_0 * e_0 = e_0 \quad . &(3.21b)\cr}$$
We see then that the 4 bosonic elements $e_0,\ \phi_1 ,\ \phi_2$, and 
$\phi_3$ define the usual quaternion algebra. Therefore, Eqs. (3.21) may 
be regarded as a super generalization of the quaternion algebra. However, 
it is no longer associative when the product involves fermionic element. 
We can moreover prove that the algebra is super-quadratic, super-flexible, 
and super-Lie-admissible, although we will not go into detail.

We can repeat a similar analysis for octonion algebra. We now consider a 
system consisting of $j=0,\ j=3,$ and $j = {3 \over 2}$. For products 
involving $j = {3 \over 2}$, the corresponding Clebsch-Gordan algebra will 
be given by
$$\psi_{3 \over 2} (m_1) \cdot \psi_{3 \over 2} (m_2) = a_1 \sum_M C
\pmatrix{{3 \over 2} &{3 \over 2} &3 \cr
\noalign{\vskip 4pt}%
m_1 &m_2 &M\cr} \psi_3 (M) \eqno(3.22a)$$
$$\eqalign{\psi_{3 \over 2} (m_1) \cdot \psi_3 (M_1) &= -\psi_3 (M_1) 
\cdot \psi_{3 \over 2} (m_1)\cr
&= a_2 \sum_{m_2} C \pmatrix{{3 \over 2} &3 &{3 \over 2}\cr
\noalign{\vskip 4pt}%
m_1 &M_1 &m_2\cr} \psi_{3 \over 2} (m_2)\cr} \eqno(3.22b)$$
for some constants $a_1$ and $a_2$. Then, a similar construction gives the 
octonion algebra for the bosonic space, and the algebra may be considered 
also as a super generalization of the octonion algebra. However, we will 
not go into its detail.

\medskip

\noindent {\bf 4. \underbar{Quantum Clebsch-Gordan Algebra}}

\medskip

The idea explained in the previous sections can be extended for any quantum 
group $L$. Consider the quantum group su$_{\rm q}$(2) which is 
defined$^{4)}$ by the commutation relations
$$\eqalignno{[H,J_\pm] &= \pm J_\pm &(4.1a)\cr
\noalign{\vskip 4pt}%
[J_+ , J_-] &= {t^{2H} - t^{-2H} \over t - t^{-1}} &(4.1b)\cr}$$
for a constant parameter $t (= q^{1 \over 2})$. The co-product $\Delta\ :\  L 
\rightarrow L \otimes L$ is specified by
$$\eqalignno{\Delta (J_\pm) &= t^{-H} \otimes J_\pm + J_\pm \otimes t^H 
\quad , &(4.2a)\cr
\Delta (H) &= 1 \otimes H + H \otimes 1 &(4.2b)\cr}$$
which satisfies
$$\Delta ([x,y]) = [\Delta (x), \Delta (y)]\quad . \eqno(4.3)$$
Moreover, the anti-pode $S \ :\ L \rightarrow L$ operates as
$$S (t^{\pm 2H}) = t^{\mp 2H} \quad , \quad S(J_\pm) = - t^{\pm 2} J_\pm 
\eqno(4.4)$$
which obeys anti-morphism relation
$$S (xy) = S(y) S(x) \quad . \eqno(4.5)$$
Finally, the co-unit $\epsilon$ is given by
$$\epsilon (t^{\pm 2H}) = 1 \quad , \quad \epsilon (J_\pm) = 0 \quad . \eqno
(4.6)$$
These operations define the Hopf algebra, i.e.
$$\eqalignno{(\Delta \otimes id) \circ \Delta &= (id \otimes \Delta ) 
\circ \Delta \quad , &(4.7a)\cr
( \varepsilon \otimes id) \circ \Delta &= (id \otimes \epsilon ) 
\circ \Delta = id \quad , &(4.7b)\cr
\epsilon \circ S &= \epsilon \quad , &(4.7c)\cr
\sigma \circ (S \otimes S) \circ \Delta &= \Delta \circ S &(4.7d)\cr}$$
where $\sigma$ in Eq. (4.7d) stands for the permutation operation.

Let $|j,m>_q$ now be the representation of su$_{\rm q}$(2) 
with$^{4)}$
$$\eqalignno{H |j, m>_q &= m|j,m>_q &(4.8a)\cr
J_\pm |j,m>_q &= ([j \mp m ]_q [j\pm m+1]_q)^{1 \over 2} |j,m \pm 1>_q
&(4.8b)\cr}$$
where
$$[n]_q = {t^n - t^{-n} \over t-t^{-1}} = t^{(n-1)} +
t^{(n-3)} + \dots + t^{-(n-1)} \eqno(4.9)$$
for non-negative integer $n$. Then, the quantum Clebsch-Gordan algebra for 
the integer angular momentum state $j$ will be given by
$$|j,m_1>_q \cdot |j,m_2>_q\  = b_j \sum_{m_3} C_q \pmatrix{j &j &j\cr
\noalign{\vskip 4pt}%
m_1 &m_2 &m_3\cr}
|j ,m_3>_q \eqno(4.10)$$
for the quantum Clebsch-Gordan coefficient$^{4),7)}$
 $C_q \pmatrix{j_1 &j_2 &j_3\cr
m_1 &m_2 &m_3\cr}$.  This replaces Eq. (1.1). The product defined by Eq. 
(4.10) behaves covariantly under actions of su$_{\rm q}$(2) in the following 
sense. Let $m\ :\ V \otimes V \rightarrow V$ be the multiplication 
operation in the $2j+1$ dimensional representation space $V$, i.e.
$$m (x \otimes y) = x \cdot y \quad , \quad x,\ y\ \epsilon\ V \quad .
\eqno(4.11a)$$
Then, by the construction of the quantum Clebsch-Gordan coefficients, it 
must satisfy the relation
$$g \circ m = m \circ \Delta (g) \eqno(4.11b)$$
for any $g\ \epsilon$ su$_{\rm q}$(2). In order to illustrate that Eq. 
(4.11b) is the statement of covariance, let us consider the case of the 
ordinary su(2) Lie algebra corresponding to the choice $t=1$. Then,
$$\Delta (g) = g \otimes 1 + 1 \otimes g$$
so that Eq. (4.11b) operated to $x \otimes y$ will reproduce the standard 
formula
$$g(x \cdot y) = (gx) \cdot y + x \cdot (gy)$$
for the action of the Lie algebra as a derivation.

In what follows, we will restrict ourselves to the special case of $j=1$ 
and set 
$$\eqalignno{x_0 &= |1 ,0>_q \quad , &(4.12a)\cr
x_\pm &= |1, \pm1 >_q \quad . &(4.12b)\cr}$$
Then, Eq. (4.10) will lead to the multiplication table of
$$\eqalignno{x_0 \cdot x_0 &= \beta (t-t^{-1}) x_0 \quad , &(4.13a)\cr
x_0 \cdot x_\pm &= \mp \beta t^{\mp 1} x_\pm \quad , &(4.13b)\cr  
x_\pm \cdot x_0 &= \pm \beta t^{\pm 1} x_\pm \quad , &(4.13c)\cr  
x_\pm  \cdot x_\mp &= \pm \beta  x_0 \quad , &(4.13d)\cr  
x_\pm \cdot x_\pm &= 0  &(4.13e)\cr}$$
for a suitable normalization constant $\beta$, which satisfies Eq. (4.11b). 
For $t=1$, this reproduces the results of section 2. Note that the algebra 
given by Eqs. (4.13) is no longer anti-commutative. It still possesses a 
involution operation $\omega$ defined by
$$\omega (x_0) = x_0 \quad , \quad \omega (x_\pm) = x_\mp \eqno(4.14)$$
which satisfies
$$\omega (x \cdot y) = \omega (y) \cdot \omega (x) \quad . \eqno(4.15)$$

For simplicity, we will normalize the constant $\beta$ to be $\beta =1$ in 
what follows. Then, the associator given by
$$(x,y,z) = (x \cdot y)\cdot z - x \cdot (y\cdot z) \eqno(4.16)$$
can be verified to satisfy
$$(x,y,z) = B (x,y) z - B(y,z)x \eqno(4.17)$$
where the bilinear form $B(x,y)$ is defined by
$$\eqalign{&B(x_+ ,x_-) = t \quad , \quad B(x_-, x_+) = {1 \over t} \quad
,\cr
&B (x_0, x_0) = -1 \quad , \quad B(x_\pm , x_0) = B(x_0 , x_\pm) =0 \quad .
\cr}\eqno(4.18)$$
As a matter of fact, we have
$$B( x_{m_1}, x_{m_2}) =\ {\rm constant}\ C_q \pmatrix{ 1 &1 &0\cr
\noalign{\vskip 4pt}%
m_1 &m_2 &0\cr} \quad . \eqno(4.19)$$
We also note that $B(x,y)$ is no longer symmetric but is associative, i.e.
$$B (x \cdot y, z) = B(x, y\cdot z) \quad . \eqno(4.20)$$
Moreover, it satisfies
$$B(\omega (x), \omega (y)) = B(y, x) \eqno(4.21)$$
for the involution $\omega$ given by Eq. (4.14).

From Eq. (4.17), we see that the algebra is not flexible, but is
 Lie-admissible$^{1)}$ since it obeys
$$(x,y,z) + (y,z,x) + (z,x,y) = 0 \quad . \eqno(4.22)$$

If we now adjoin the unit element $e_0$ with the new product * by
$$x * y = x \cdot y - B(x,y) e_0 \quad , \eqno(4.23)$$
it is easy to verify from these equations that it is associative, i.e.
$$(x*y) *z = x* (y*z) \quad . \eqno(4.24)$$
Actually, the new algebra is isomorphic to the quaternion algebra so that 
ththe quantum quaternion algebra is nothing but the same as the usual
quaternion algebra.

We can apply the same method for systems involving both $j=1$ and ${1 \over 
2}$ states to obtain a
 quantum-deformed super algebra of osp(1,2). 
Analogously, if we consider $j=3$, then it will lead to a quantum 
generalization of the octonion algebra. However, by a reason not given 
here, we have to actually use the quantum deformation of the 7-dimensional 
representation of the exceptional Lie algebra $G_2$ rather than the $j=3$ 
states of su(2) in order to properly describe the quantum octonion algebra. 
These, however, will be studied in the future.  

\medskip

\noindent {\bf \underbar{Acknowledgement}}

\medskip

The present article is dedicated to the 90th birthday of Professor Ta-You 
Wu of Tsin-Hua and Jiao Tong Universities. It is also supported in part by 
the U.S. Department of Energy Grant No. DE-FG02-91ER40685.

\vfill\eject

\noindent {\bf \underbar{References}}

\medskip

\item{1.} S. Okubo, {\it Introduction to Octonion and Other 
Non-Associative Algebras in Physics} (Cambridge University Press, 
Cambridge, 1995).

\item{2.} H.C. Myung, {\it Malcev-Admissible Algebras} (Birkh\"auser, 
Boston, 1986).

\item{3.} M. Scheunert, {\it The Theory of Lie Super Algebras} 
(Springer-Verlag, Berlin, 1979).

\item{4.} L.C. Biedenharn and M.A. Lohe, {\it Quantum Group Symmetry and 
q-Tensor Algebras} (World Scientific, Singapore, 1995).

\item{5.} H. Weyl, {\it The Classical Groups} (Princeton University, 
Princeton 1939).

\item{6.} S. Okubo, Jour. Math. Phys. {\bf 35}, 2785 (1994).

\item{7.} A.N. Kirillov and N.Yu. Reshitikhin, USSR Academy of Sciences
 (unpublished) 1988, L. Vaksman, Sov. Math. Dokl. {\bf 39}, 467 
(1989).

\end